\documentstyle[12pt]{article}
\begin{document}
\title{Multilayer Thermionic Refrigerator and Generator}
\author{G.D. Mahan, J.O. Sofo \cite{cab}, and M. Bartkowiak\\[0.1in]
Department of Physics and Astronomy\\ University of Tennessee, Knoxville,
TN,37996-1200, and\\[0.1in]
Solid State Division\\ Oak Ridge National Laboratory\\ Oak Ridge, TN 37831-6030}
\date{\today}
\maketitle
\begin{center}Abstract\end{center}

A new method of refrigeration is proposed. Cooling is obtained by thermionic emission of
electrons over periodic barriers in a multilayer geometry. These could be either Schottky
barriers between metals and semiconductors or else barriers in a semiconductor superlattice. The
same device is an efficient power generator. 
A  complete  theory is provided.
\newpage
A new method is proposed for refrigeration and power generation. The devices are composed of
multilayers of periodic barriers. The currents are perpendicular to the barriers. Such devices
are presently under consideration for thermoelectric cooling and power generation. Here we
propose that the same devices can be used for efficient thermionic cooling, or power generation.

 What is the difference between a thermoelectric device\cite{97pt,98ssp} and a thermionic
one\cite{94gdm,97shak}? Their descriptions are remarkably similar. In both cases one puts
a temperature gradient on a semiconductor. Both devices are based upon the idea
that electron motion is electricity. But the electron motion also carries energy. Forcing a
current transports energy for both thermionic and thermoelectric device. The basic difference
seems to be whether the current flow is ballistic or diffusive. In thermionic motion, the device
has relatively high efficiency if the electrons ballistically go over and across the barrier.
They carry all of their kinetic energy from one electrode to the other. In thermoelectric
devices, the motion of electrons is quasiequilibrium and diffusive. One can describe the energy
transport by a Seebeck coefficient, which is an equilibrium parameter. There have also been
recent theories of nonequilibrium thermoelectric effects\cite{97zak}. They seem to be midway
between the thermionic(ballistic) and thermoelectic(quasiequilibrium) regimes. We note that many
semiconductor superlattices, with short periods, are being made\cite{97venk} and measured
along the
$c$-axis. These periods are so short that electron motion over the barrier is probably
ballistic. We suggest that these devices cool by thermionic effects rather than thermoelectric
ones. 

Earlier we suggested\cite{94gdm} that thermionic devices could be used as refrigerators. For the
usual device of two metal plates separated by an air gap, cooling at room temperature requires a
low work function. No metals have values that low $(\phi \sim 0.3$ eV). Numerous groups have
suggested\cite{94row}-\cite{97shak} that such small barriers are easily attainable in
semiconductor systems. Here the barriers are semiconductors, while the electrodes could be
either metals or other semiconductors. We considered this geometry in our original paper, but
thought that the thermal conductivity of the solid would be a major obstacle. Here we show that
the thermal effects can be dealt with by going to a multilayer geometry. We show that such
devices have efficiencies which could be twice those of thermoelectric ones. 

 The physics
behind thermionic cooling is simple. Most physicists are familiar with the technique of cooling
liquid helium-4 by pumping the vapor from the cryostat. The most energetic helium atoms  leave
the liquid and become gas molecules. Pumping them away removes these energetic atoms, thereby
cooling the liquid. In thermionic refrigeration, one uses a voltage to sweep away the most
energetic electrons from the surface of a conductor. Those electrons with sufficient energy to
overcome the work function are taken away to the hot side of the junction. Removing the energetic
electrons from the cold side cools it. Charge neutrality is maintained at the cold side by
adding electrons adiabatically through an ohmic contact. 

Thermionic devices must have the electrons ballistically traverse the barrier in order to have a
high efficiency. This requires that the mean-free-path $\lambda$ of the electron in the barrier
be longer than the width $L$ of the barrier. This constrains the barrier width $L$ to be rather
small. This fact is a key feature of the analysis. The general constraints are
\begin{eqnarray}
\lambda &>& L > L_t\\
L_t &=&  \frac{\hbar}{2k_BT}\sqrt{\frac{e\phi}{m^*}}
\end{eqnarray}
where $L_t$ is the minimum thickness to prevent the electron from tunneling through the
barrier. Values of $\L_t$ seem to be around 5-10 nm for most semiconductors. The values of
$\lambda$ are less certain. There are good measurements of the electron mean-free-path for
motion along the layers, but little data for motion perpendicular to the layers. If these
values are similar, then one can easily find that $\lambda \sim$ 50-100 nm. Thus there is room
in the above inequality to find values of $L$ which work. 

The general form of the energy currents for a single barrier are
\begin{eqnarray}
J_Q &=& J_{Qe} - \frac{\delta T}{R_1}\\
R_1 &=& 2 R_I + \frac{L}{K} + \frac{L_e}{K_e}\label{5.2}
\end{eqnarray}
The first term in the energy current is the electron part from thermionic emission. It is
given below. The second term is the phonon part, which contains the thermal resistance $R_1$
for one barrier. The thermal resistance depends upon the thickness $(L_e)$ and thermal
conductivity $(K_e)$ of the electrodes, as well as two interface ('Kapitza') terms
$R_I$\cite{pohl,89sch}. The largest term will usually be $L/K$ for the semiconductor barrier. The
problem is that $L$ is small which makes $R_1$ small which makes $\delta T/R_1$ big. This is the
problem with the thermal conductivity. It can be overcome by having $\delta T$ be small. This is
the reason for the multilayer geometry. Each barrier can support only a small temperature
difference
$\delta T_i$. A macroscopic temperature $\Delta T $ is obtained by having $N$ layers so that
$\Delta T = N\langle \delta T_i\rangle$. Most of our modeling assumes that $ \delta T_i$ has an
average value of 1-2 $^{\circ}$C. Below we show that adequate cooling power is available for
these values.
\section{Refrigeration}
\subsection{Single Barrier}
The first step in the derivation is to solve for the currents over a single barrier of width $L$.
 We assume the barrier is a constant at
zero applied voltage, which means it has a square shape. At a nonzero voltage $\delta V$ the
barrier  has the shape of a trapezoid.  The formulas for the
electrical  $(J)$ and heat $(J_Q)$  currents are given in terms of the hot
and cold temperatures $(T_h, T_c)$\cite{94gdm}
\begin{eqnarray}
J_{Rj} &=& AT_j^2e^{-e\phi/k_BT_j}\label{2}\\
A &=& \frac{emk_B^2}{2\pi^2\hbar^3}{\cal T}\label{3}\\
J &=& J_{Rc}-J_{Rh}e^{-e\delta V/k_BT_h}\label{4}\\
eJ_Q &=& [e\phi+2k_BT_c]J_{Rc}\nonumber\\ &-& [e\phi+2k_BT_h]J_{Rh}e^{-e\delta V/k_BT_h}
-\frac{\delta T}{R_1}\label{5}\\
\delta T &=& T_h - T_c
\end{eqnarray}
These equations are for a single barrier. 
Eqn.(\ref{2}) is the standard Richardson's equation\cite{sze} for the thermionic current
over a work function $e\phi$ which in this case is the Schottky barrier height
between the metal and semiconductor. Alternately, it is the barrier in a semiconductor
quantum well.  The factor of
${\cal T}$ denotes the fraction of electrons transmitted from the metal  to the semiconductor.
It is calculated using quantum mechanical matching of the wave functions.  The formulas for $J$
and $J_Q$ assume the bias $e\delta V$ is to lower the Fermi level on the hot side, so
that the net flow of electrons is from cold to hot.

For electrons the charge $e$ is negative which makes $\phi,J$ also negative. We find this
confusing to treat, so we take $e$ and $\phi$ as positive as if the system were a hole
conductor. Then $J>0$ for particle flow to the right, which we are assuming.

We show that for a single layer the optimal value of applied bias $e\delta V
\propto \delta T$ which is also small. Denote as $T$ the mean temperature of
the layer, and then $T_c = T-\delta T/2, T_h = T+\delta T/2$. Then we expand
the above formulas for the currents in the small quantities $(\delta T/T, e\delta V/k_BT)$
and find, after some algebra
\begin{eqnarray}
J &=& \frac{eJ_R}{k_BT}[\delta V - V_J]\label{6}\\
J_Q &=& J_R(b+2)[\delta V - V_Q]\label{6.1}\\
b &=& \frac{e\phi}{k_BT}\\
eV_J &=& k_B\delta T[b+2]\\
eV_Q &=& k_B\delta T q\label{7}\\
q &=& b+2 + u\\
u &=& \frac{2+Z}{b+2}\\
Z &=& \frac{e}{k_BR_1J_R}=Z_0e^b\\
Z_0 &=& \frac{ek_B}{R_1A(k_BT)^2{\cal T}}= \left(\frac{T_R}{T}\right)^2\\
(k_BT_R)^2 &=& \frac{2\pi^2\hbar^3}{m k_B R_1 {\cal T} }
\end{eqnarray} 
The symbol $J_R$ denotes Richardson's current eqn.(\ref{2}) at temperature
$T$. The dimensionless constant $Z_0$  plays an
important role in the results. Refrigeration requires that
$\delta V >V_Q$. High efficiency requires that $Z_0<1$. We have also introduced the dimensional
constant $T_R$ which is determined by the thermal resistance and the effective mass.

The efficiency $\eta$ of   a single barrier is defined as the heat flow divided by
the power input
\begin{eqnarray}
\eta &=& \frac{J_Q}{\delta VJ}\label{8}\\
&=& \frac{k_BT(b+2)}{e}\frac{\delta V-V_Q}{\delta V(\delta V-V_J)}
\end{eqnarray}
We now vary $\delta V$ in the above equation to find the value
of $\delta V=V_m$ which gives the maximum efficiency $\eta_m$ which is
\begin{eqnarray}
V_m &=& V_Q + \sqrt{V_Q(V_Q-V_J)}\\
\eta_m &=& \left(\frac{T}{\delta T}\right)\hat{\eta}\label{10}\\
\hat{\eta}(\phi) &=& \frac{b+2}{(\sqrt{q}+\sqrt{u})^2}\label{11}
\end{eqnarray}
 The first factor on the right in
eqn.(\ref{10}) is approximately the Carnot efficiency of the one-layer system. The remaining
factor
$\hat{\eta}$ is the reduction of this efficiency due to the inefficiency of the device. The
latter factor depends upon the parameter
$b=e\phi/k_BT$ and  the dimensionless factor $Z_0$  contains the
thermal resistivity $R_1$. Fig. 1 
shows the function $\hat{\eta}(\phi)$ for several values of $T_R$ in the range
between 100 and 500 K. It is important to have $\hat{\eta}$ be as large as possible.
Clearly this constraint limits the value of $T_R$ to smaller values, which has $Z_0$ to be 
less than one.  For larger values of
$Z_0$ the effective efficiency becomes too small to be useful.

The value of $T_R$ is reduced by the Kapitza resistance $R_I$ at the interfaces.  Numerical
estimates for $R_I$ are given by\cite{pohl,89sch}. The device will only be efficient when $R_I$
large. Large thermal boundary resistance can be achieved by having acoustic
mismatch between the metal and semiconductor layers. The important point is that the thermal
conductivity term has to stay small. The thermal resistivity is large in semiconductor
superlattices\cite{93spr,95yu,96chen,96cap,97lee,97per}.

The results for the single barrier demonstrate that one has to have small values of barrier
height $e\phi \approx 2k_BT$ and large values of the thermal resistance. These characteristics
carry over to the theory of the multilayer device. Small values of the barrier heights are
reported in several references\cite{97venk,73Kaji,78Lera,96hicks}
\subsection{Multiple Barriers}
Now  consider  the multilayer thermionic refrigerator. There are $N$ barriers,
with alternating electrodes. Thus there are $2N+1$ layers, assuming electrodes come
first and last. Denote
$\delta T_i$ and
$\delta V_i$ as the temperature and voltage change across one barrier. These values change
 from the initial to the final layer. This change is due to the generation and flow
of heat. At each barrier, the electron ballistically crosses the barrier region, and then loses
an amount of energy
$e\delta V_i$ in the electrode. The heat generated at each electrode must flow out of the
sample according to the equation
\begin{eqnarray}
\frac{d}{dx}(J_{Qi}) &=& J\frac{\delta V_i}{L_i}\label{20}
\end{eqnarray}
where $L_i$ is the effective width one one barrier plus one metal electrode.  We take
$L_i$ to be a constant, although it could vary with $i$. The width of the device is $D=NL_i$. In normal
usage there would be a negative sign on the right-hand side of this equation. It is absent
since we changed the sign of $J$. 

 The functions change slowly with $i$ and we  treat them as
continuous variables
$\delta T_i/L_i=dT/dx$ and $\delta V_i/L_i=dV/dx$. The variation of  $\delta V_i$ upon 
$i$ is unknown, but we guess that it is proportional to $\delta T_i$. Introduce the
dimensionless function $v(x)$
\begin{eqnarray}
e\delta V_i &=& k_B\delta T_i v\label{21}\\
J &=& J_R \frac{\delta T_i}{T}[v-b-2]\label{22}\\
eJ_Q&=& J_R(b+2) k_B\delta T_i[v-q]\label{23}
\end{eqnarray}
We use eqn.(\ref{22}) to define $\delta T_i$ since the current $J$ is the only constant
among all of the parameters
\begin{eqnarray}
\delta T &=& \frac{JT}{uJ_R}I\label{24}\\
I &=& \frac{u}{v-b-2}\label{25}\\
D\frac{dT}{dx} &=&\frac{ITyZ}{u}\\
y &=& J\left(\frac{k_B}{e}\right)R_N\\
R_N &=& N R_1.
\end{eqnarray}
The variable $I(x)$ is introduced in eqn.(\ref{24},\ref{25}). After trying various alternatives,
we decided it is the  convenient variable for the present problem.  The parameter $y$ is a
dimensionless current. Eventually we vary this parameter to find the optimal efficiency. The
thermal resistance of the entire device is denoted as
$R_N$. Its value is irrelevant since it ocurs in $y$ which is a variational parameter. Use 
eqn.(\ref{24}) to evaluate $\delta V_i$ in (\ref{21}) and  this result is inserted into
eqn.(\ref{20})
\begin{eqnarray}
J_Q &=& J(\phi + 2\frac{k_B T}{e})(1-I)\\
D \frac{d}{dx}(I)&=& -y\frac{Z}{2+Z}[u + I(b+2I)]\label{28}\\
D\frac{dT}{dx} &=& \frac{yITZ}{u}\label{29}
\end{eqnarray}
Eqns.(\ref{28}) and (\ref{29}) describe the variation in temperature
and voltage across the multilayer device. The input parameters are the barrier height
$e\phi (b = e\phi/k_BT)$ and the  parameter $T_R$ which  depends on the thermal resistance.
The parameter
$y$ is variational. One solves the equations for different values of
$y$ and chooses the value which give the best results. One selects an initial guess for
$I(x=0)=I_c$ at the cold end along with $T(0)=T_c$. Then one can iterate the above two
equations using one sided derivatives. The value of $I_c$ is varied until one finds the
desired value of $T_h$. The calculation is done for different values of $y$ until the maximum
efficiency is attained. The Carnot efficiency is defined as the heat taken from the cold side
$J_{Qc}$ divided by the input power $J\Delta V$
\begin{eqnarray}
\eta &=& \frac{J_{Qc}}{J \Delta V}\label{30}\\
J_{Qc} &=& J\left(\frac{k_BT_c}{e}\right)(b_c+2)[1-I_c]\\
\Delta V &=& \sum_{i=1}^N\delta V_i =
\left(\frac{yk_B}{eN}\right)\sum_i\frac{T_iZ(T_i)}{u_i}[u+I_i(b_i+2)]
\end{eqnarray}
where $\Delta V$ is the net voltage drop across the multilayer device. The above integral for
$\Delta V$ is easy after one has solved for $T(x)$ and $I(x)$.

Fig.2 shows the efficiency of a refrigerator, with $T_c$=260 K and $T_h$= 300 K,  for
 four values of $T_R = 100-500 K$. They are shown as a function of 
of $\phi$. The efficiencies are quite high: $\eta$=2.5
for
$T_R$ = 100 K to $\eta$=0.6 for $T_R$ = 500 K. The comparable value for the thermoelectric
refrigerator, with a dimensionless figure of merit equal to unity, is $\eta_{TE}$ = 0.70. That
is comparable to the thermionic results when $T_R$ = 500 K. Thus one wants to operate the
thermionic device with values of $T_R$ less than 500 K. 

 The factor of
$T_R$ is determined by the thermal resistivity. Reducing the thermal resistivity by a factor of
25 only reduces the efficiency by a factor of 5. Thus the efficiency scales with $T_R$ rather
than with $R_1$. This is quite different than in thermoelectric devices, where the result is
almost directly proportional to the thermal resistivity. Note in eqn.(\ref{28}) that
$Z$ enters in the formula in the combination of $Z/(2+Z)$ which saturates at large values of $Z$.
The increase is $Z_0$ is offset by a lower value of barrier height $\phi$.

The exact  results show that the thermionic refrigerator has about 30\% of the
ideal Carnot efficiency $\Delta T/T_c$ for values of $T_R$ = 200-300 K. Our calculations show
that the efficiency  slowly declines with increasing values of $T_R$. The value of $\phi$
where the maximum efficiency occurs also declines with increasing value of $T_R$. 
\section{Multilayer Thermionic Power Generation}
Power generation can be understood by considering a simple device with one 
barrier between two electrodes. If the electrodes are at different temperatures, and if there
is no initial voltage between the electrodes, then electrons are thermally excited over the
barrier. A net electron flow goes from hot to cold. If the electrodes are insulated, then the
electrodes become charged and the system develops an open circuit voltage, which opposes the flow
of electrons and reduces it to zero. Connecting the electrodes to an external circuit causes
current to flow. Power can be extracted from the device. The behavior is identical in concept to
a solar cell. Here the efficiency is calculated using the usual definition: it is the external
power $J\Delta V$ divided by the heat extracted from the hot electrode $J_{Qh}$. 
\subsection{One Barrier}
Again the barriers
are assumed to be thinner than the mean-free-path of the electrons, so that one can apply the
formulas of thermionic emission. There is a small temperature drop $\delta T$ and voltage
drop $\delta V$ across the barrier. The efficiency of a single barrier using
eqns.(\ref{6})--(\ref{7}) is
\begin{eqnarray}
\eta &=& \frac{e}{k_B T(b+2)}\frac{\delta V(\delta V-V_J)}{\delta V-V_Q}
\end{eqnarray}
The voltage $\delta V$ is varied to find the value  $\delta V_m$ which gives the maximum
efficiency $\eta_m$
\begin{eqnarray}
\delta V_m &=& \frac{K_B\delta T}{e}\sqrt{q}(\sqrt{q}-\sqrt{u})\\
\eta_m &=& \frac{\delta T}{T}\hat{\eta}
\end{eqnarray}
where $\hat{\eta}$ is defined in eqn.(\ref{11}). It is the same function which is found in the
efficiency of the refrigerator. 
\subsection{N-Barriers}
Again assume that the device has $2N+1$ layers. Barriers alternate with electrodes, with
electrodes on both ends. The electrodes can be metals, or else conducting semiconductors. The
barriers are  going to be nonconducting semiconductors. Again we shall find that the
barrier height $e\phi$ is going to be small. There is a small temperature drop
$\delta T_i$ and voltage drop $\delta V_i$ across each barrier $i$. Eqns.(\ref{6})-(\ref{7})
also apply to the generator. The difference from the refrigerator is that now $J<0$ and $J_Q<0$.
This makes the parameter $y$ be negative. Another change is that $b+2>v$ so that the function
$I(x)$ is negative. Thus we can apply eqns.(\ref{28},\ref{29}) with the understanding that $y$
and $I(x)$ are both negative.

The formula for the efficiency is the inverse of
the one for the refrigerator
\begin{eqnarray}
\eta &=& \frac{e\Delta V}{(e\phi + 2k_BT_h)(1-I_h)}
\end{eqnarray}
These equations were solved on the computer. We  assumed that $T_c$ = 300 K and $T_h$ = 400 K, so
that $\Delta T$ = 100 K. Figure 3 shows the efficiency as a function of barrier
height for several values of $T_R$. This should be compared with a thermoelectric
generator which has an efficiency of $\eta$=0.048 for the same operating
temperatures. Clearly the thermionic device is more efficient for small values of $T_R$.
Figure 4 shows the  temperature and $\delta V_i$ profiles along the thermionic device, assuming
that $N=100$.
\section{Thermoelectric Analogy}
Thermionic devices are not thermoelectric devices. However, it is useful to examine the analogy
between the two kinds of  devices. Once we linearize the eqns. (\ref{6},\ref{6.1},\ref{20}) for
small voltage drops, and then make them continuous in a multilayer geometry, they become
identical in form to the equations of a thermoelectric. This analogy gives the effective
formulas for  the conductivity
$\sigma$, the Seebeck coefficient $S$, and the thermal conductivity $K$. 
\begin{eqnarray}
\sigma &=& \frac{e J_R L_i}{k_B T}\label{te1}\\
S &=& \frac{k_B}{e}(b+2)\label{te2}\\
K &=& [\frac{k_B }{e}J_R + \frac{1}{R_1}]L_i \label{te3}\\
{\cal Z} &=& \frac{\sigma S^2 T}{K} = \frac{b+2}{u}\label{te4}
\end{eqnarray}
The first term in the thermal conductivity is the electronic contribution $K_e$. Note that it
obeys a modified Wiedemann-Franz Law
\begin{eqnarray}
K_e &=& \sigma T \left(\frac{k_B}{e}\right)^2
\end{eqnarray}
 The last line in eqn. (\ref{te4}) gives the dimensionless figure of
merit ${\cal Z}$. It is known in thermoelectric devices that the maximum efficiency is obtained
by having 
${\cal Z}$  be as large as possible. Varying $b$ to maximize ${\cal Z}$ gives the relationship 
\begin{eqnarray}
be^b &=& \frac{4}{Z_0} = \frac{2mk_B R_1}{\pi^2\hbar^3(k_B\bar{T})^2}
\end{eqnarray}
where $\bar{T}$ is the mean temperature in the device.
This is an accurate estimate of the optimal barrier height $\phi$, where $b=e\phi/k_B\bar{T}$.
Furthermore, the thermoelectric estimates of the efficiency of the refrigerator $(\eta_r)$ and
generator $(\eta_g)$ 
\begin{eqnarray}
\eta_r &=& \frac{\gamma T_c-T_h}{\Delta T(\gamma +1)}\label{te5}\\
\eta_g &=& \frac{\Delta T(\gamma -1)}{\gamma T_h + T_c}\label{te6}\\
\gamma &=& \sqrt{1 + {\cal Z}}
\end{eqnarray}
are remarkably accurate compared to the computer solutions. The Eqs.\ (\ref{te5}) and ({\ref{te6}) estimate the correct efficiency with an accuracy of about 1\%.
These formulas provide analytical estimates of the efficiency. One can also use the
thermoelectric analogy to find the current densities at maximum efficiency as well as other
parameters.

 In thermoelectric devices it is
highly desirable but rather rare to have values of ${\cal Z}>1$. However,
in modeling thermionic devices we find values of ${\cal Z}$ much larger than one for reasonable
values of thermal resistance. The effective Seebeck coefficient in eqn.(\ref{te2}) is about
250-300 $\mu$V/K since values of $b$ are between one and two. By using the low values of
thermal conductivity reported along the $c$-axis of a superlattice, we estimate that $T_R$ =
200-400 K. For this range of parameters the effective dimensionless figure of merit ${\cal Z}$ is
between 2 and 5. The efficiencies of the thermionic devices are correspondingly much higher than
in thermoelectric devices. Ballistic transport carries more heat than diffusive flow. 

\section{Discussion}
A detailed theory is presented of the properties of a multilayer thermionic refrigerator and
power generator. It is showns that if the thermal resistance is high, that the devices can be
twice the efficiency of the equivalent thermoelectric devices. The values of thermal
resistiviity needed to make them work well are in the range of reported values for multiple
quantum wells. It appears these devices are practical.

One of the interesting questions is to select the value of $N$. This determines the number of
multilayers. As long as this number is larger than about ten in refrigerators, it does not 
change the efficiency. The choice of $N$ may affect the thermal resistance. For a fixed
temperature drop $\Delta T$, the value of $N$ affects the average temperature drop per layer
$\langle\delta T\rangle=\Delta T/N$. The cooling power is the energy current from the cold
side. An estimate is
\begin{equation}
 J_{Qc} = \frac{A \bar{T} \phi \Delta T}{N}\exp(-e\phi/k_B \bar{T})
\end{equation}
where $\bar{T}$ is the average temperature of the device. For a refrigerator take $A$ = 120
A/(cm$^2$ K$^2$) along with $\bar{T}$ = 280 K, $\phi$ = 0.050 eV, and $\langle\delta T\rangle$ =
1K. Then the result is $J_{Qc} =$ 212 W/cm$^2$. This is a large cooling power. Thus having a
small temperature drop permits ample cooling. It does require a small barrier height.
Incidently, the numerical solutions allow an accurate calculation of $J_{Qc}$ and the above
formula is an underestimate by about a factor of two or three.  One could increase $N$ by a
factor of ten and still have ample cooling power. The manufacturing costs would depend on $N$.
Another issue is mechanical strength. A thicker device (larger $N$) is stronger. Thus there are
trade offs between cooling power, manufacturing costs,  and mechanical strength. Such
engineering issues are beyond the scope of the present discussion.

We thank L. Woods for helpful discussions.  Research support is acknowledged
from the University of Tennessee, and from Oak Ridge National Laboratory managed by Lockheed
Martin Energy Research Corp. for the U.S. Department of Energy under contract DE-AC05-96OR22464.

\subsection*{Appendix}
Here we examine the transport of electricity and heat using the drift-diffusion eqn.(1). We
will show that the amount of heat is negligibly small when one uses this equation. Unless
the layer thickness is in the regime where thermionic emission is valid, the semiconductor
barrier does not transport significant amounts of heat.

 We consider a square barrier for the semiconductor.
Assume there is a small  applied potential and a small temperature difference
$\delta T = T_h-T_c$. Then the formula for the current should be
\begin{eqnarray}
J &=& e\mu n \frac{\delta V}{L} - \mu k_BT
\left[\frac{n(L)-n(0)}{L}\right]\\
n(0) &=& n_0 e^{-e\phi/k_BT_c}\\
n(L) &=& n_0 e^{-e\phi/k_BT_h}\\
n &=&n_0 e^{-e\phi/k_BT}
\end{eqnarray}
We expand the density exponents using $T_{h,c} = T\pm \delta T/2$ and find
\begin{eqnarray}
J &=& \frac{e\mu n}{L}[\delta V - V_D]\\
V_D &=& b k_B\delta T
\end{eqnarray}
The above formula resembles eqn.(\ref{6}). The two constant voltages $V_J$ and $V_D$ are
similar. The prefactors are very different. Denote by $r$ the ratio of the two prefactors,
which gives the ratio of the magnitudes of the currents predicted by drift-diffusion
compared to the currents predicted by thermionic emission
\begin{eqnarray}
r &=& \frac{e\mu n}{e J_R/(k_BT)}\\
&=& \frac{\sigma}{\sigma_0}\left(\frac{L_0}{L}\right) e^b\\
\sigma_0 &=& \frac{e^2 m L_0 k_B T}{2\pi^2 \hbar^3}
\end{eqnarray}
where $\sigma$ is the conductivity of the semiconductor and 
 $L_0$ is a characteristic length which we take to be one micron. At room temperature we
find that $\sigma_0$ = 8.1 kS/m.  InSb has a mobility of 8 m$^2$/(V s) at $n =
10^{20}/$m$^3$ for $\sigma$ = 130 S/m. Even for $b=4$ this predicts a ratio of $r\approx
10^{-3}$. So the current due to drift-diffusion is one thousand times smaller than the
current due to thermionic emission. The energy currents have the same ratio of
proportionality. Thus the flow of heat is negligible when the thickness of the semiconductor
is greater than the mean free path of the electron, so that the
particles diffuse.

\newpage
\section*{Figure Captions}
\begin{enumerate}
\item The reduction in the Carnot efficiency $\hat{\eta}$ for a single barrier as a
function of barrier height $\phi$ for  values of $T_R$= 100, 200, 300, 400, and 500 K.
\item Efficiency $\eta$ of a multilayer thermionic refrigerator with $T_c$ = 260 K and
$T_h$ = 300 K. Results are plotted as a function of barrier height for different values of
$T_R$.
\item Exact efficiency of a multilayer thermionic power generator as a function of
barrier height
$\phi$ for several values of $T_R$. We assumed $T_c$ = 300 K and $T_h$ = 400 K.
\item Change in the temperature $T_i$ and voltage $V_i$ as a function of position along
the multilayer power generator. We assumed that $T_c$ = 300 K, $T_h$ = 400 K, $\phi$ = 0.054 V,
and $T_R$ = 200 K.
\end{enumerate}
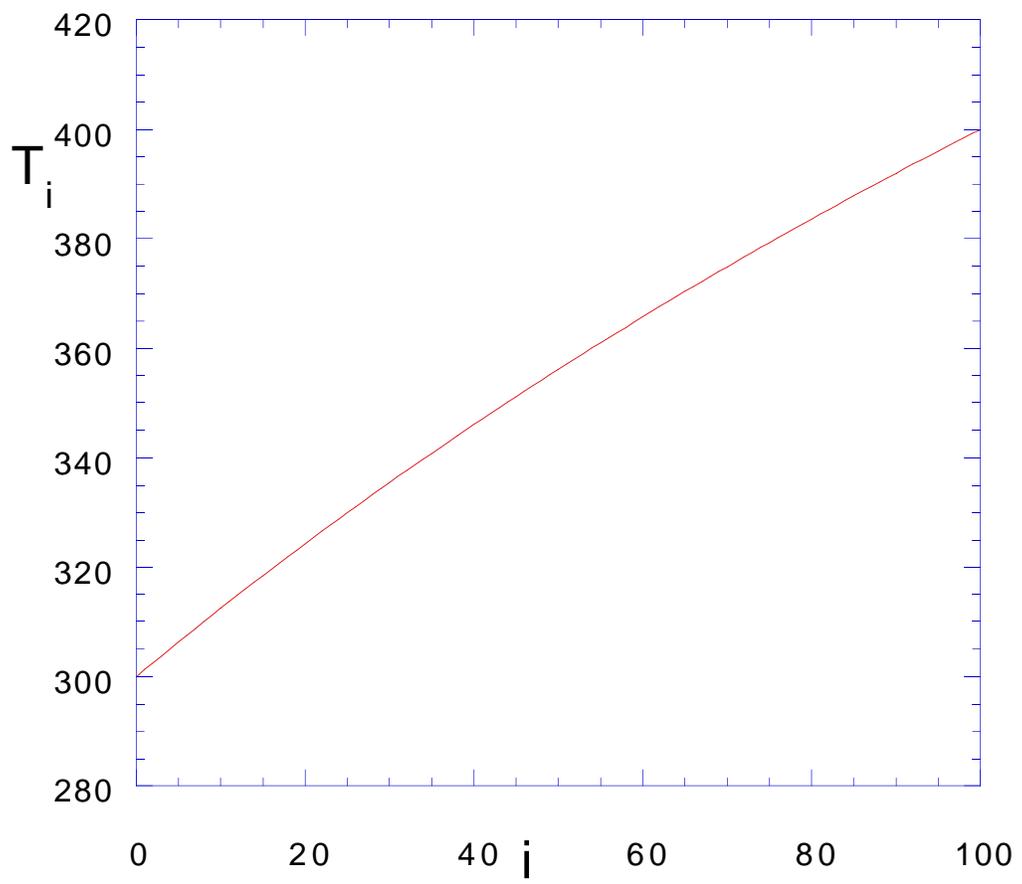
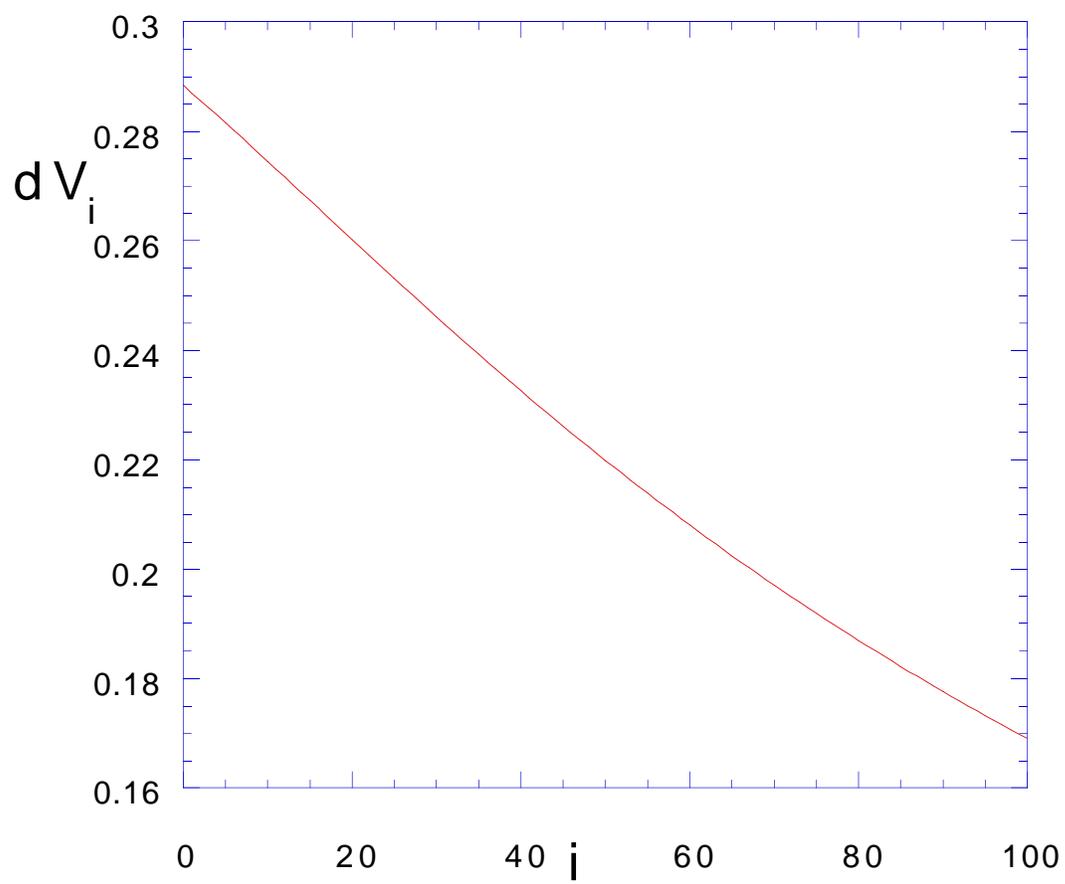
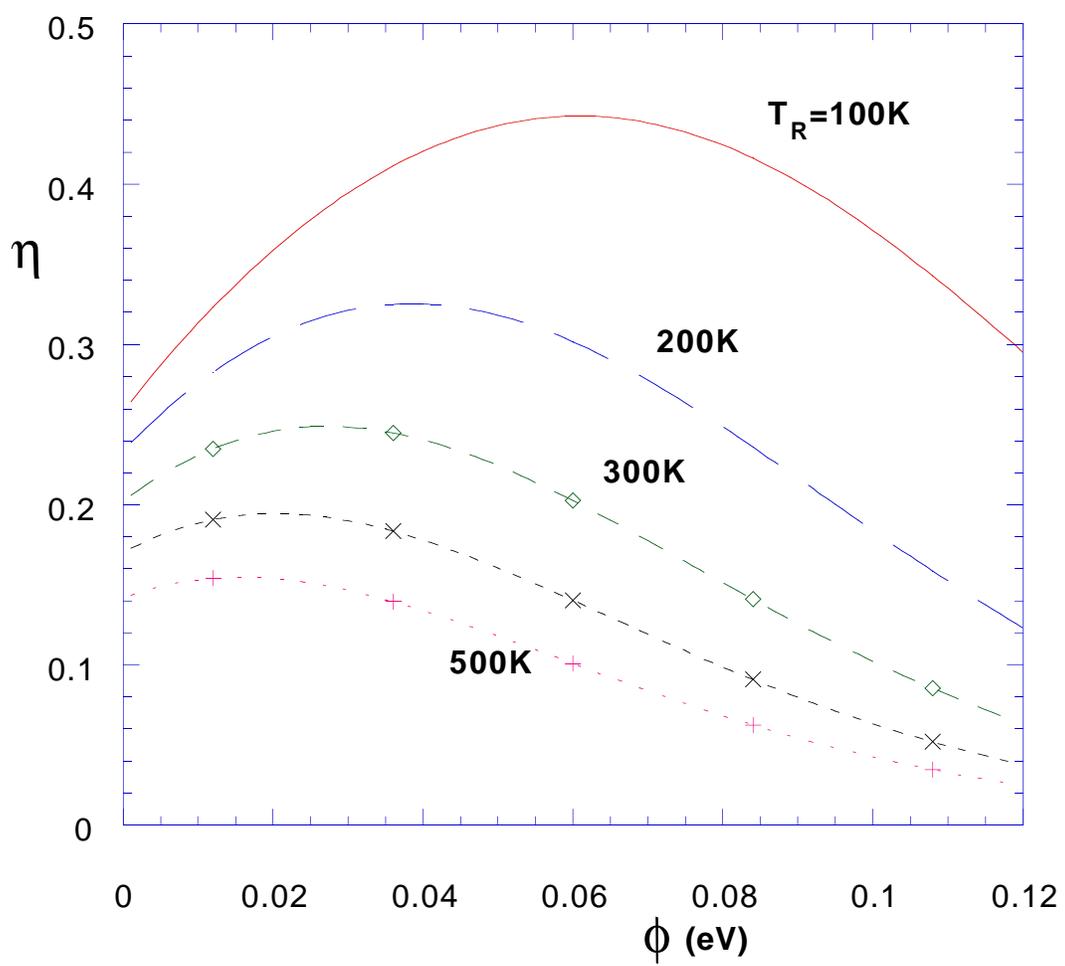
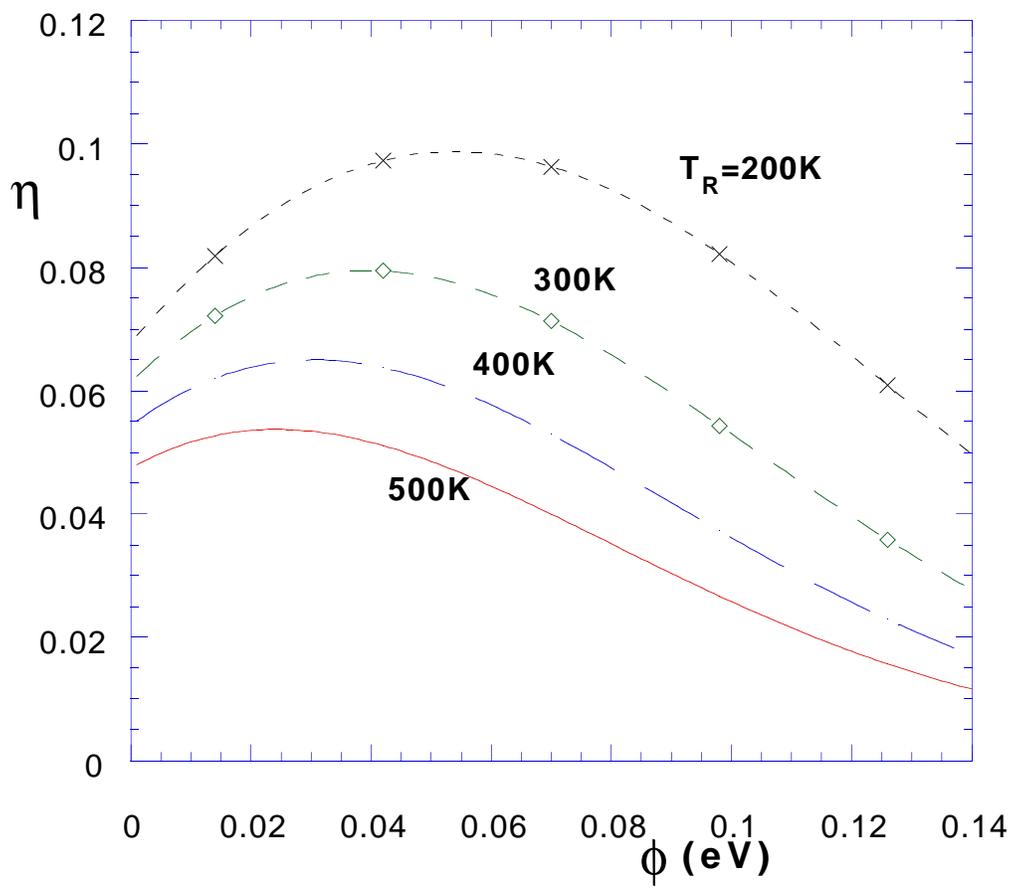
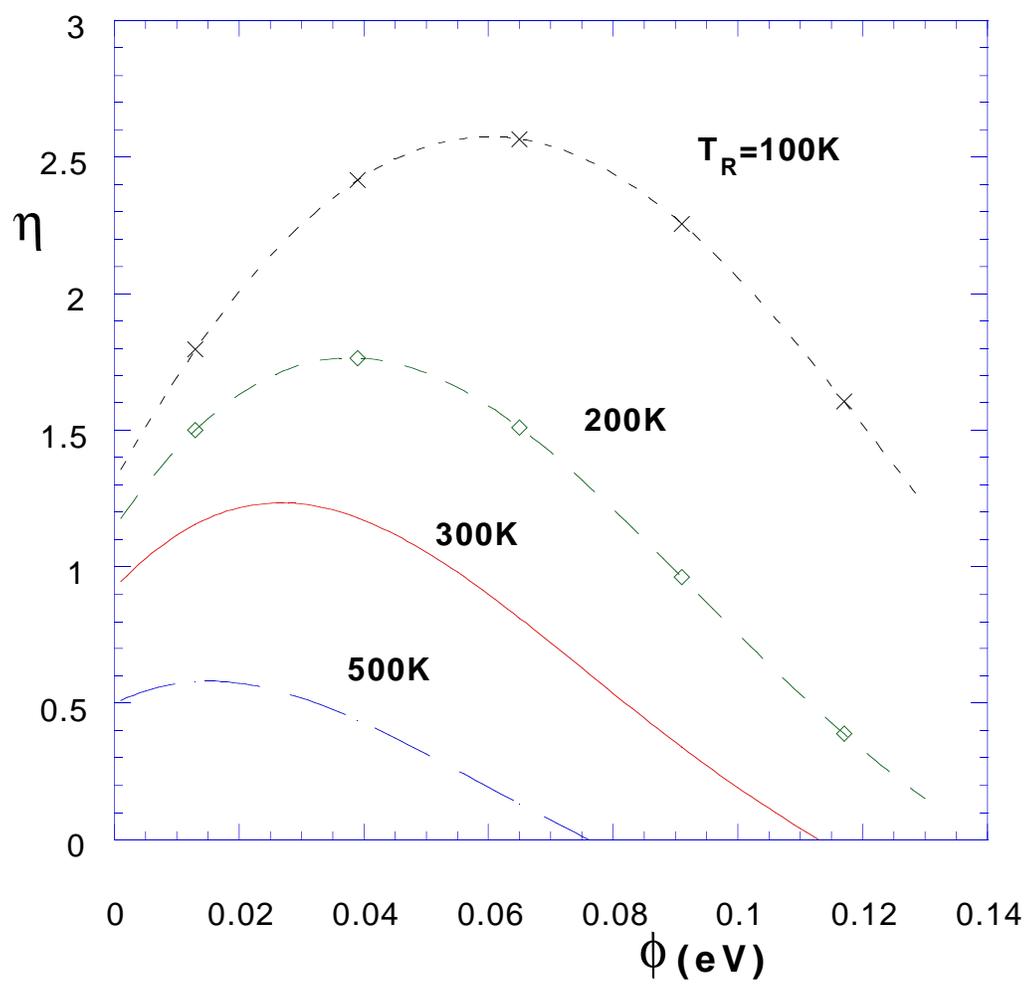
\end{document}